\documentclass[aps,pre,twocolumn,showpacs,groupedaddress,letterpaper,superscriptaddress]{revtex4-1} % for review and submission 

\usepackage{graphicx}  % needed for figures
\usepackage{epstopdf}
\usepackage{subcaption}
\usepackage{ragged2e}
\usepackage{amssymb}   % for math
\usepackage{xcolor}

\newcommand{\beq}{\begin{equation}}
\newcommand{\eeq}{\end{equation}}
\newcommand{\bea}{\begin{eqnarray}}
\newcommand{\eea}{\end{eqnarray}}

\begin{document}

\title{Investigating molecular mechanism for the stability of ternary systems containing cetrimide, fatty alcohol 
and water by using computer simulation}

% \author{Vu Dang Hoang$^{1}$, Tran Huu Hung$^{2}$, Cao Phuong Cong$^3$, Hue Minh Thi Nguyen$^{2}$, Toan T. Nguyen$^{*3}$}

% \affiliation{
% $^1$Department of Analytical Chemistry and Toxicology,
% Hanoi University of Pharmacy,\\
% 13-15 Le Thanh Tong, Hoan Kiem, Hanoi, Vietnam\\
% $^2$Faculty of Chemistry and Center for Computational Science, Hanoi National University of Education, \\
% 136 Xuan Thuy Street, Cau Giay, Hanoi, Vietnam,\\
% % $^2$Thuy Loi University, \\
% % 175 Tay Son Street, Dong Da, Hanoi, Vietnam,\\
% $^3$Faculty of Physics and VNU Key Laboratory on Multiscale Simulation of Complex Systems, \\
% VNU University of Science, Vietnam National University, \\
% 334 Nguyen Trai Street, Thanh Xuan, Hanoi, Vietnam\\
% }

\author{Vu Dang Hoang}
\affiliation{Department of Analytical Chemistry and Toxicology,
Hanoi University of Pharmacy,
13-15 Le Thanh Tong, Hoan Kiem, Hanoi, Vietnam}

\author{Hung Huu Tran}
\affiliation{Faculty of Chemistry and Center for Computational Science, Hanoi National University of Education,
136 Xuan Thuy Street, Cau Giay, Hanoi, Vietnam}

\author{Cao Cong Phuong}
\affiliation{Faculty of Physics and VNU Key Laboratory on Multiscale Simulation of Complex Systems, VNU University of Science, Vietnam National University, 334 Nguyen Trai Street, Thanh Xuan, Hanoi, Vietnam}

\author{Hue Minh Thi Nguyen}
\affiliation{Faculty of Chemistry and Center for Computational Science, Hanoi National University of Education,
136 Xuan Thuy Street, Cau Giay, Hanoi, Vietnam}

\author{Toan T. Nguyen}
\email[Corresponding author: ]{toannt@hus.edu.vn}
\affiliation{Faculty of Physics and VNU Key Laboratory on Multiscale Simulation of Complex Systems, VNU University of Science, Vietnam National University, 334 Nguyen Trai Street, Thanh Xuan, Hanoi, Vietnam}

% \date{\today}

\begin{abstract}
Computer simulations using atomistic model are carried out to investigate the stability of ternary systems of pure or mixed fatty alcohols, cetrimide, and water. These semi$-$solid oil-in-water systems are used as the main component of pharmaceutical creams. Experiments show that the mixed alcohol systems are more stable than pure ones. The current experimental hypothesis is that this is the result of the length mismatch of the alkyl chains. This leads to higher configurational entropy of the chain tip of the longer alcohol molecules. Our simulation results support this hypothesis. The results also show that the shorter alcohol molecules become stiffer with higher values of the deuterium order parameters and smaller area per molecule. The magnitude in fluctuations in the area per molecule also increases in mixed systems, indicating a higher configurational entropy. Analysis of the molecular structure of simulated systems also shows good agreements with experimental data.
\end{abstract}

% all pacs numbers can be found in https://ufn.ru/en/pacs/all/
% 87.10.Tf Molecular dynamics simulation
% 87.15.N- Properties of solutions of macromolecules 
% 87.15.He Dynamics and conformational changes
% 81.16.Dn Self-assembly
% 87.16.A- Theory, modeling, and simulations 
% 87.16.D− Membranes, bilayers, and vesicles
% 87.18.Bb Computer simulation
% 87.19.rm Structure 
% 88.20.fj Mixed alcohols
% 82.70.Uv Surfactants, micellar solutions, vesicles, lamellae, amphiphilic systems,

% \keyword{MD, mem, alcoholic}

\maketitle

\section{Introduction}

Pharmaceutical creams are one of the common types of topical formulations, used as a
means of delivering an active ingredient directly to the skin. Generally speaking, they
are oil-in-water semisolid emulsions.

\begin{figure}[t!]
    \centering
    \includegraphics[width=7cm]{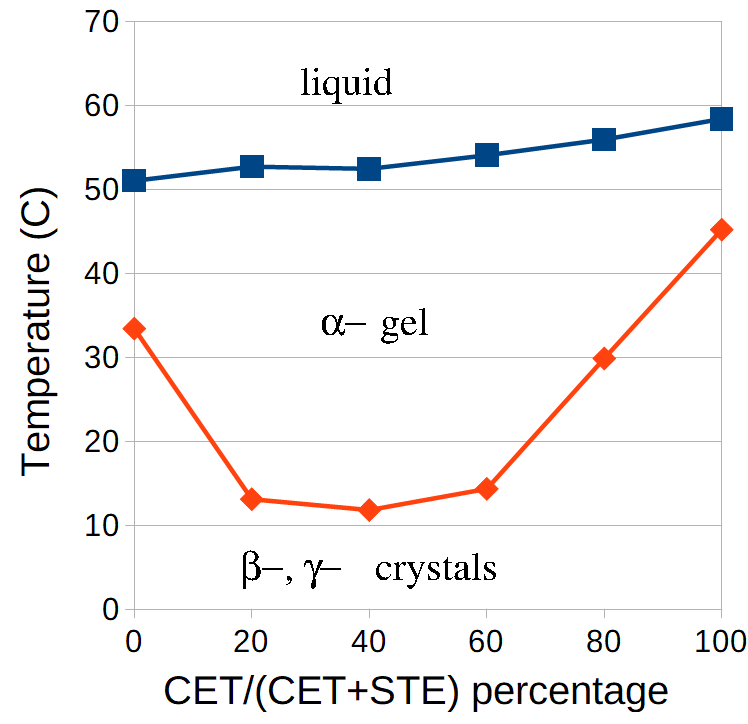}
    \caption{(Color online) The phase diagram of the mixture of stearyl
and cetyl alcohol at various temperature and mixing ratio. The
$\alpha-$gel (liquid crystal) exists to relatively low temperature
while for pure cetyl or stearyl bilayers, the $\mathbf{L}_\beta$ 
crystal phase exists at higher temperature \cite{Iwata2017}.
}
\label{fig:phase}
\end{figure}

According to the gel network theory \cite{Eccleston1997, junginger1984}, these
formulations consist of at least four phases: 
(i) crystalline/hydrophilic gel phase is
composed of bilayers of surfactant and fatty amphiphile and water molecules are
inserted between the bilayers (i.e. interlamellarly fixed water); 
(ii) water molecules
bound as free bulk water are in equilibrium with the interlamellarly fixed water in the gel
phase; 
(iii) lipophilic gel phase is built up by the excess of the fatty amphiphile, 
which is
not part of the hydrophilic gel phase; and 
(iv) dispersed oil phase (i.e. inner phase) is
mainly immobilized mechanically from the lipophilic gel phase. 
In practice, surfactant-fatty alcohol-water ternary systems could be used as representative
models of the continuous phases of the corresponding semisolid o/w emulsions.
Previous works on the characterization of semisolid oil-in-water emulsions containing
pure and mixed homologue fatty alcohols (i.e. cetyl, stearyl and cetostearyl alcohols)
and cetrimide \cite{Iwata2017, Eccleston1985, Eccleston1977} have shown that the oil-in-water emulsions
increased in consistency and remained stable on the addition of cetostearyl alcohol,
whereas the emulsion stability decreased markedly to mobile liquid when cetyl or stearyl
alcohol was added individually on the preparation. This is demonstrated
in the phase diagram, Fig. \ref{fig:phase}, where it is shown that
the $\alpha-$gel is stable down to lower temperature for 
mixed cetyl and stearyl alcohol as compared to either
of them individually. 

Although the above-mentioned studies partially showed 
the correlation between various fatty alcohol combinations and
their influence on the microstructure and consistency of 
the corresponding oil-in-water emulsions prepared with them, 
no comprehensive studies at the molecular level has been
carried out. The aim of this study is to elucidate the molecular mechanism 
for the higher stability of the mixed cetostearyl bilayer 
$\alpha-$gel network as compared to each component used separately.
Previous experiments \cite{Snyder1983, Iwata2017} suggested 
this is due to the alkyl chain length mismatch in the mixed system. 
As a result, the alkyl chain tip of C18OH has a high degree of freedom, and, 
in this part {\em gauche} coordination increases. 
Our computational study supports this hypothesis at the molecular level.
We also show that upon mixing the alkyl chain
of C16OH becomes stiffer and the C18OH alkyl chain tip
becomes more flexible. Also, the in$-$plane area per molecule
fluctuations is more pronounced for mixed systems. This explanation provides
further insights into the molecular mechanism of
the stability of mixed system.

The paper is organized as follows. After the Introduction section,
in Sec. \ref{sec:model}, the model of our system
and various physical parameters used in the simulation are presented in details.
In Sec. \ref{sec:res}, the
results are presented and  their relevance to 
available experimental data is discussed.
We conclude in Sec. \ref{sec:conclusion}.

\section{Computational details}
\label{sec:model}

The chemical structure of cetyl (CH$_3$(CH$_2$)$_{15}$OH), 
stearyl (CH$_3$(CH$_2$)$_{17}$OH) alcohols
and cetrimide (C$_{17}$H$_{38}$BrN) are downloaded
from pubchem database (https://pubchem.ncbi.nlm.nih.gov/)
with CID 2682, 8221 and 14250 respectively.
Gaussian 09 software package is used to optimize
their three dimensional structures, and to calculate the 
partial charges on each atom. These partial charges are then adjusted
and rounded to match the values of the hydrocarbon tails,
hydroxyl and choline head groups of forcefield for the united atom model for polyalcohols \cite{AUA4ff} and of the slipid forcefields
\cite{Jaembeck2012, Jaembeck2013}. These parameters are then
combined with  the bonded interactions in the Amber GAFF forcefield by
using AmberTools 2017 to create the full forcefield for the molecules
used in the computer simulations. For the bromide ions,
the AMBER99-ILDN forcefield is used. For the water molecules,
the common TIP3P forcefield is used for its high compatibility
with the AMBER forcefields.

The initial coordinates of the atoms of a bilayer of these molecules are 
then constructed 
in the following way. First, the corresponding optimized atom coordinates
in the previous step are duplicated at random site in either a square or hexagonal
lattice in the $xy$-plane with a lattice constant of 0.48nm.
The hydrocarbon tails are directed inward the bilayer along the
$z$-axis, while the hydroxyl polar group are exposed outward the solution
(see Fig. \ref{fig:startConf}).
We prepare the system of cetyl and stearyl alcohol mixture at 100:0, 70:30, 
50:50, 30:70, 0:100 number ratios. For easy identification in later
discussion, we name these systems 100CET, 70CET, 50CET, 30CET, and 0CET
respectively. Beside the alcohol molecules,
there are also positively charged cetrimide molecules 
embedded in the bilayer to make them stable. Their number is 
about 10\% of the total number to match experimental systems.
The bilayer is then solvated with 37210 water molecules, 
creating simulation box of about 24nm along the $z$-axis.  
The electrostatic repulsion between parallel bilayer 
among these cetrimide molecules keeps the bilayer structure stable. 
To compensate for the charges of cetrimide,
the system is made neutral by replacing randomly 64 water molecules
by bromide ions, Br$^{-}$. 

The periodic boundary condition is used for all three directions. The thickness
of the water layer (24 nm) has been chosen so that it is much larger than
the typical electrostatic screening in the system (about 1nm) and
the electrostatic coupling between neighboring simulation boxes is well
screened out. This value also lies in the range of the repeat distance of these
cetyl/stearyl lamellar value from 6 to 100nm \cite{Eccleston2000} 
while keeping the computational cost reasonable.
For reference, the number of each type of molecules for each of
our simulated systems are listed in Table. \ref{table:number}.
Particle Mesh Ewald method \cite{darden93PME} 
was used to treat the long-range electrostatic interaction 
with a real space cutoff of 1.4 nm. Van der Waals interactions are also cut off at 1.4 nm, 
with the appropriate cut-off corrections added to pressure and energy.
\begin{center}
\begin{table}
\begin{tabular}{l|r|r|r|r|r}
\hline\hline
System name  & 0CET & 30CET & 50CET & 70CET & 100CET \\
\hline
STE:CET \# ratio  & 100:0& 70:30 & 50:50 & 30:70 & 0:100 \\
\hline
\# of Cetyl &	0 &176 & 292 & 408 &584 \\
\# of Stearyl &	584& 408 & 292 & 176 & 0\\
\# of Cetrimide &64 & 64 & 64 & 64 &64\\
\# of Br ions &64 & 64 & 64 & 64 &64\\
\# of water molecules &	37146 & 37146 & 37146 & 37146 & 37146 \\
\hline
\end{tabular}
\caption{The five systems simulated in this work with different number
ratios of stearyl:cetyl alcohol molecules varying from 0 to 100\%. 
The name of the system 
listed will be used throughout this paper for easy identification.
The last five rows list the number of molecules for different species of molecules of the simulated systems. }
\label{table:number}
\end{table}
\end{center}

After constructing the initial coordinates of the atoms and  topology of the molecules,
the systems are further processed and simulated by using GROMACS 2018.3 
molecular dynamics software package \cite{berendsen1995gromacs,Abraham2015}.
The whole system of surfactants, water and ions are equilibrated for 100 ns at temperature 300K and 
pressure of 1 atm by using NPT ensemble using Berendsen thermostat
and barostat. The barostat maintains independently
the pressure along each of the three directions $x$, $y$ and $z$ axes of the system.
This ensures that each dimension of the simulation box varies independently
from each other, allowing the bilayer molecules to relax into its optimal lattice structure
regardless of the initial lattice configuration.
For production run, the systems are simulated for additional 1$\mu s$
 at the same temperature and pressure by using Nose$-$Hoover thermostat 
and Parrinello$-$Rahman barostat.  All barostats are kept at 1 atm so that on average 
the surface tension of the bilayers simulated are zero. 
For taking statistics, all the configurations within the first 100ns of simulation,
where the systems are relaxing to its equilibrium state,
are dropped. All the simulations are performed on the High Performance
Computing cluster of the VNU Key Laboratory for Multiscale simulation of
complex systems at the Vietnam National University $-$ Hanoi. For analysis, beside
standard tools from GROMACS software, the VMD software package \cite{Humphrey1996} 
and MEMBPLUGIN plugin \cite{Guixa-Gonzalez2014} 
is used for visualization, calculation of the order parameters, area
per molecules, tilt angles, and other physical properties of the bilayer.

\section{Results and discussions}
\label{sec:res}

Experimentally, it is established by various synchrotron X$-$ray studies that
the lamellar bilayer of the cetyl, stearyl alcohols is of hexagonal lattice packing with
diffraction spacing of 4.06\AA\ for the amount of water and temperature
that we simulate  \cite{Eccleston2000,Iwata2017,Fukushima1976,Fukushima1977}. 
This corresponds to the hexagonal lattice with
lattice constant (or nearest neighbor distance) of $4.06\mbox{\AA}\times 2/\sqrt{3}=4.69$\AA\ and the area per
molecule of 19.13\AA$^2$. Additionally, in experimental studies of on the bilayer 
made of different types of molecules with different chain lengths and head groups, 
it is shown that this in$-$plane lattice distance depends very weakly on the length of the 
hydrocarbon chain, and mostly depends on the interaction among the head groups
\cite{Langmuir1917}. Therefore, our immediate concerns are to reproduce
these experimental facts to verify that our forcefield parameters are reasonable.

\subsection{Relaxation to equilibrium}

\begin{figure*}[ht]
    \centering
\includegraphics[width=0.90\textwidth]{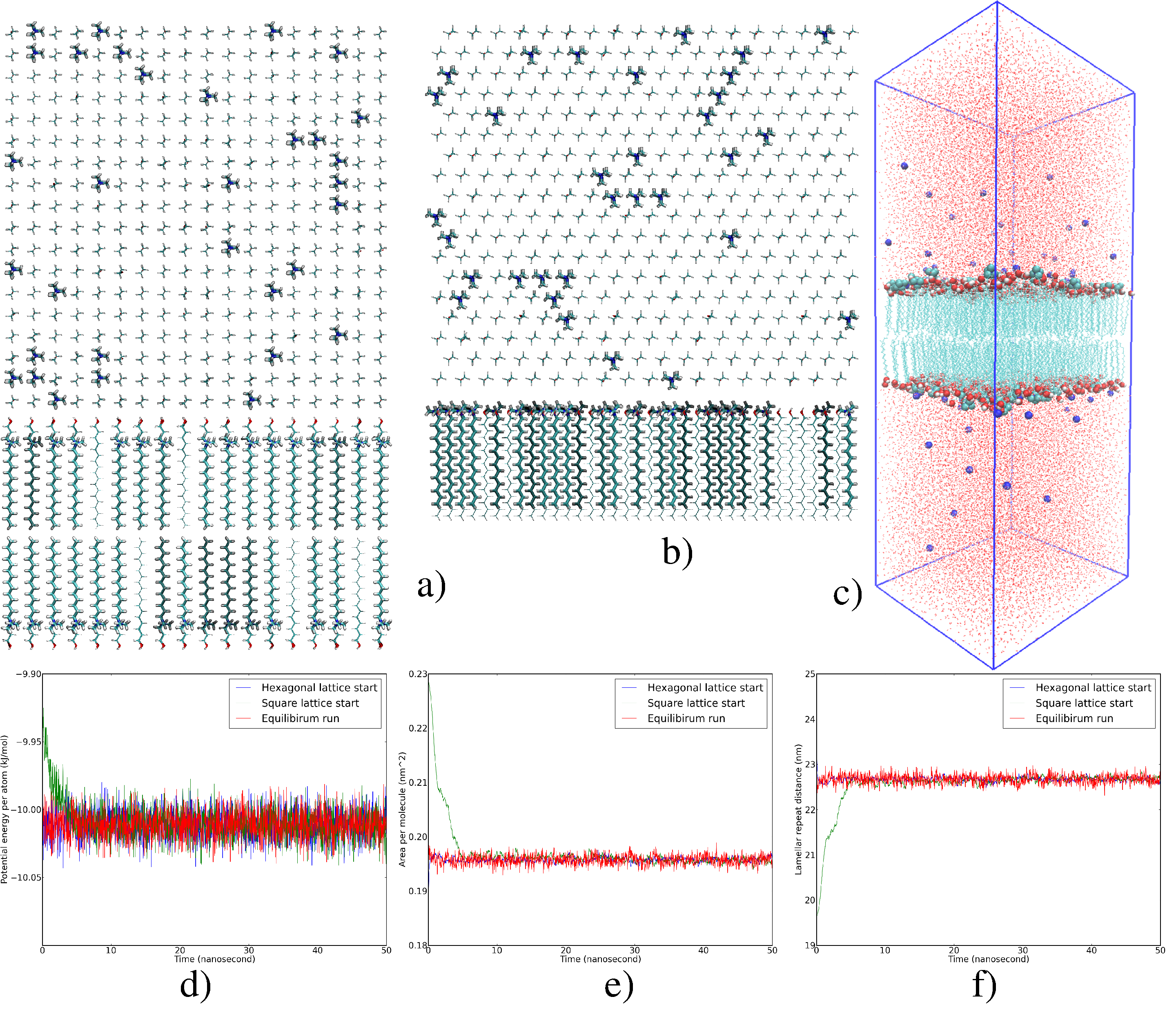}
\caption{(Color online) (a) Top view and side view of the square lattice starting configuration
of the bilayer of alcohol molecules.  The hydrocarbon tails of the molecules are aligned along the z axis
with the polar groups exposed to the outside of the bilayer.
In the $xy$ plane, they are arranged into a square lattice of lattice constant
0.48nm. The cetrimide molecules are drawed using thicker lines for clarity.
(b) Top and side views of the hexagonal lattice starting configuration of the
bilayer.
(c) A typical snapshot of our simulation system during simulation showing 
the bilayer membrane fluctuates in explicit water molecules (pink) and 
neutralizing bromide ions (blue).
The periodic simulation box is shown using by blue lines. 
(d) The relaxation of the potential energy for different starting configurations.
The same quantities obtained from a well-equilibriated system is plotted
for comparison. (e) and (f) similar to (d) but the quantities plotted are
the in$-$plane projected are per molecule and the lamellar repeat distance.
}
\label{fig:startConf}
\end{figure*}

The first question one would like to address is whether our systems have sufficient time
within our simulation to relax into its equilibrium structure, at least 
at short range order. 
To do this, two very different initial starting configurations of the bilayer is used
(see Fig. \ref{fig:startConf}). In the first starting configuration, the 292 alcohol
molecules and 32 cetrimide molecules in each leaflet are positioned randomly
at the sites of a 18$\times$18 square lattice with a lattice constant of 0.48nm 
(Fig. \ref{fig:startConf}a). The initial simulation box dimension
in the in$-$plane direction is 8.46mm$\times$8.46nm. 
In the second starting configuration, these 324 molecules are positioned
at the sites of a 18$\times$18 hexagonal lattice with a lattice constant of 0.46nm
(Fig. \ref{fig:startConf}b). The initial simulation box dimension in the in$-$plane 
direction is 8.42nm$\times$7.17nm. These two initial systems are both solvated 
with the same numbers of water molecules and Br$^-$ ions as listed in Tab. \ref{table:number}. 
They are then relaxed in NPT ensemble of 1 atm and 300$^{\circ}$K by using
Berendsen thermostat and barostat. Each dimension of the simulation box
can relax independently of each other, hence the bilayer can adopt to any optimal
gel or crystal packing that minimizes the system free energy. In Fig. \ref{fig:startConf}d, e and
f, we plot as a function of time, the potential energy of the system per atom,
the in$-$plane projected area per molecule, and the repeat distance of the
lamellar bilayer (the $L_z$ dimension of the box). For comparison, these
quantities are also plotted for an equilibrium MD run after the systems has been
equilibriated, using Nose$-$Hoover thermostat and Parrinello$-$Rahman 
barostat. As one can clearly sees, after at most 20ns, all our quantities converge
to their equilibrium value regardless of initial starting configuration. 
The convergence is rather remarkable based on the fact that the individual $xy$ dimensions 
of the box, $L_x$ and $L_y$, remain very different for different starting configurations, 
but the in$-$plane area per molecule
$$\sigma = L_x \times L_y / 324, $$
is the same for all simulations within statistical errors. Although, 
these graph are plotted for the 0CET system, data for all other four systems (not plotted)
show similar behaviours. All the relevant parameters of the system converged
after about 20ns of relaxation. Additionally, as we will see in the next section,
the local order for the alcohol molecules in the bilayer is indeed hexagonal packing
with lattice spacing of 4.69\AA\ in excellent agreement with high angle X$-$ray diffraction data.

In light of these results, for the rest of this paper, unless explicitly stated,
all systems are initialized very closed to their equilibrium structure by
using hexagonal lattice starting configuration to minimize the relaxation time
of each system. Furthermore, in each analysis for equilibrium properties, data from
the first 100ns of a MD production run is dropped to make sure the systems are well equilibriated.

\subsection{Molecular structure of the bilayers}

\begin{figure*}[t!]
    \centering
        \begin{subfigure}[t]{0.45\textwidth}
        \centering
        \includegraphics[height=2in]{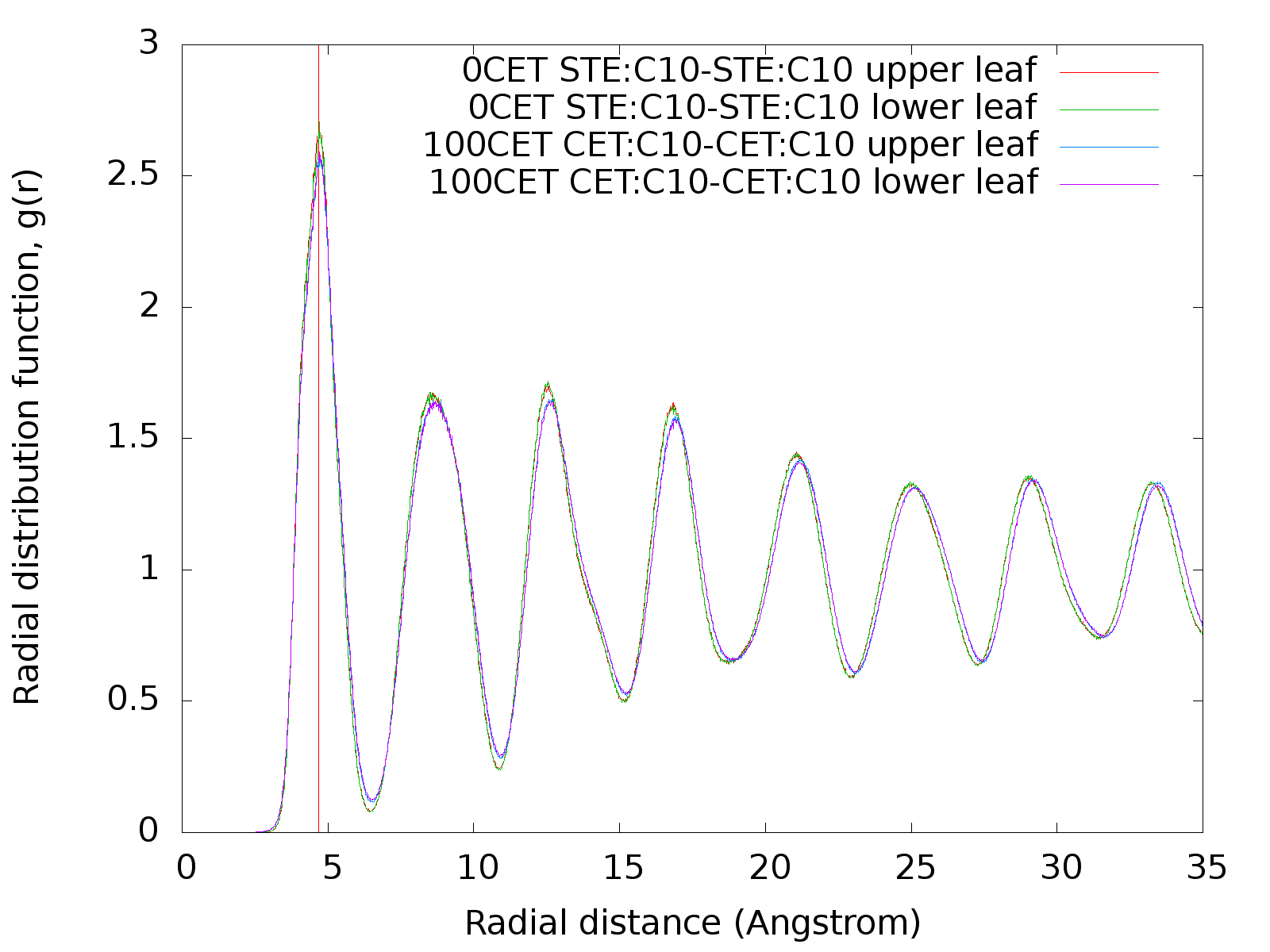}
        \caption{}
    \end{subfigure}
	~
    \begin{subfigure}[t]{0.45\textwidth}
        \centering
        \includegraphics[height=2in]{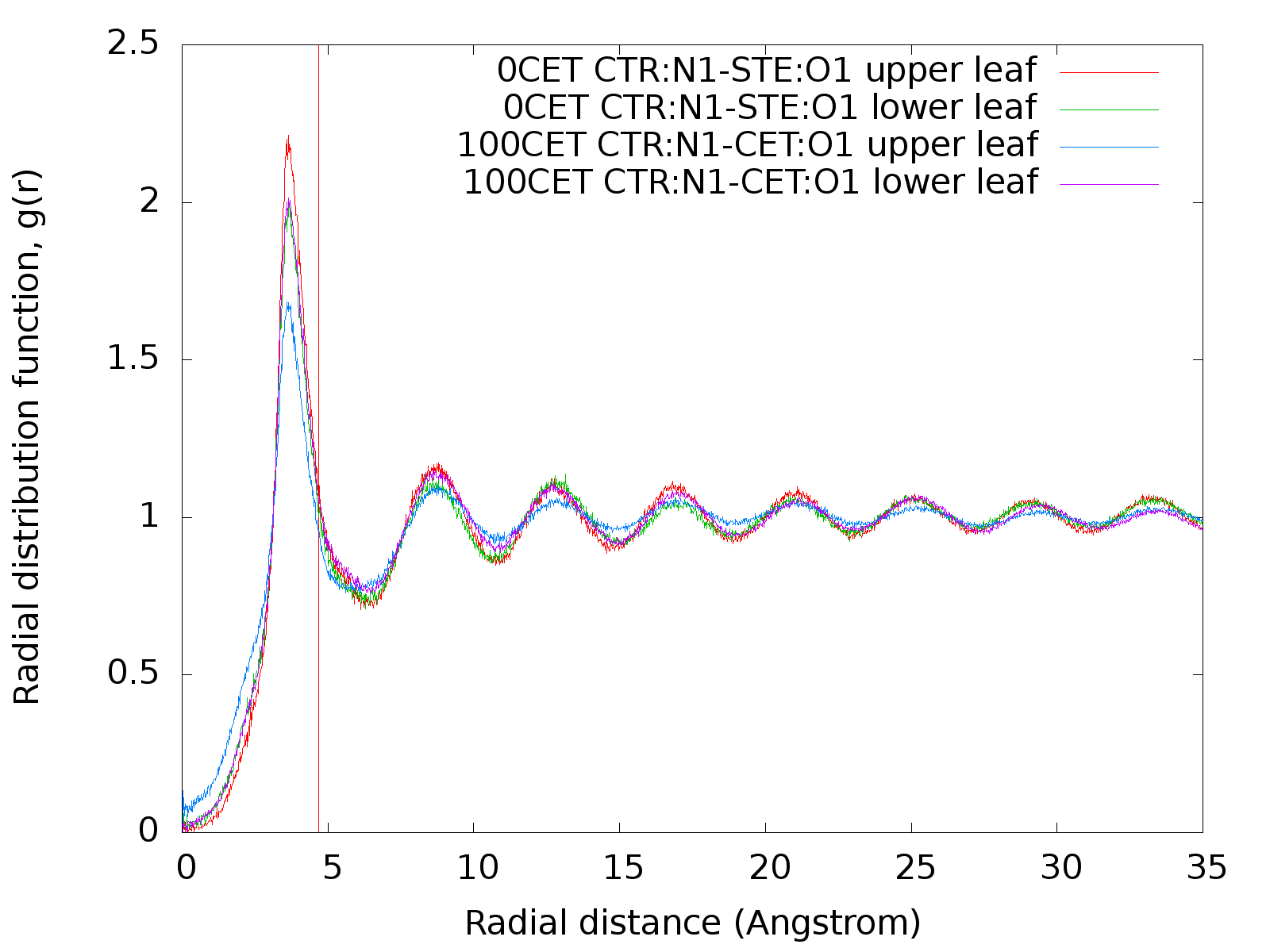}
        \caption{}
    \end{subfigure}
    \caption{(Color online) (a) The in$-$plane $xy$ radial distribution of the alcohol molecules
 calculated based on the position of the 10th carbon
 atom (counting from the $-$OH head group) of the bilayers of 0CET and 100CET systems. 
In each system, the upper and lower leaflets of the bilayer
are calculated separately. 
	(b) The in$-$plane $xy$ radial distribution function of
the oxygen atom of the polar ($-$OH) head group of the alcohol molecules 
around the nitrogen atom of the positively charged (N(CH$_3$)$_3$) head group of the cetrimide.
The red vertical line is at $r=4.69$\AA\ in both a) and b) subfigures.
}
\label{fig:rdf_0_100}
\end{figure*}

Let us turn now to describe the molecular structure of the bilayer to show
that they reproduce well the available experimental data as well as
provide additional insights obtained from the simulations. First,
the in$-$plane radial distribution function, $g(r)$,  of the carbon atoms
in the hydrocarbon tail of the molecules are calculated.
This is the probability of an atom being at the radial distance from a given atom. 
Only the in$-$plane $xy$ components of the atom positions are used for this
calculation. In Fig. \ref{fig:rdf_0_100}a, the radial distribution for the tenth carbon atom 
(counted from the OH polar head) deep in the leaflets are plotted for the systems with 100\% cetyl alcohol and 100\% stearyl alcohol. As one can sees, the first two peaks of these
distribution functions occur at almost the same locations regardless of the length
of the hydrocarbon chain. This also means that the nearest neighbor distance
between the alcohols is nearly independent of the length of the hydrocarbon chains
for the two lengths simulated. This suggests that  the hydrocarbon tail plays
small roles in determining the projected area per molecules of these bilayers.
Rather, it would be the head group (both are OH group for our molecules) that
determine the position of the peaks, as supported by experimental studies
\cite{Langmuir1917}.

Although, depending on the specific lipids, the lamella phases can be orthorhombic,
monoclinic or hexagonal crystals, the cetyl and stearyl mixture 
is known to exhibit hexagonal order in lamella phases: $\alpha-$gel
at higher temperature, and $\mathbf{L}_\beta$ crystals at lower 
temperature. X$-$ray experiments \cite{Eccleston2000} shows
hexagonal order with the diffraction spacing of 4.06\AA\ for 
various cetostearyl alcohol ternary gel. This translates into the nearest neighbor distance
of 4.69\AA\  and an unit cell area of 19.13\AA$^2$. 
Our simulation agrees well with these experimental facts.
Even though the starting configuration is that of either a square lattice
or hexagonal lattice of lattice constant 4.8\AA, both
systems settle to an equilibrium hexagonal order.
Figure \ref{fig:rdf_0_100}a shows that
the most probable nearest neighbor distance of our alcohol molecules 
projected on the $xy$ plane is $a=4.69$\AA\ in excellent match to
experimental data. Furthermore, both cetyl and stearyl molecules
show almost the same inter$-$atom distance. This suggests that
the length of the alkyl chain has only a small influence 
on the short range order in the bilayer. The difference only shows
up in a slight difference in the locations of higher order
peaks in the radial distribution. This also agrees with
experimental data \cite{Langmuir1917} showing this length depends mostly on
the type of the head group, not on the length of the alkyl chain.
% In a hexagonal lattice,
%this would corresponds to the are of the unit cell to be
%%
%\begin{equation}
%\sigma = \frac{\sqrt{3}}{2} a^2 = 0.183 \mbox{nm}^2.
%\end{equation}
%%
%In Table \ref{table:dimension}, 
%the mean dimensions of the simulation box, $L_{x,y,z}$ as well
%as the area per molecule, $\sigma=L_x\times L_y/324$ 
%obtained from our simulation are listed.
%All the systems show that this area matches well with a hexagonal
%ordering and does not depend on the composition of the bilayer
%within the uncertainty error. Further more, the average values of
%18.6-18.7$\AA^2$  for the area per molecule falls within 
%2\% of the experimental value. 
This provides strong support
for the chosen physical parameters of the atoms and molecules
that we use. Further detail investigation of these parameters
will be published in a near future work, where the phase
diagram and energetics of these alcohol systems are computed computationally.
\begin{center}
\begin{table}
\begin{tabular}{l|c|c|c|c}
\hline 
System & $L_x$(nm) & $L_y$(nm) & $L_z$(nm) & $\sigma$(nm$^2$) \\
\hline
0CET     & 8.50$\pm$0.06 & 7.38$\pm$0.05 & 23.38$\pm$0.09 & 0.194$\pm$0.003  \\
30CET   & 8.51$\pm$0.05 & 7.40$\pm$0.04 & 23.17$\pm$0.09 & 0.194$\pm$0.003 \\
50CET   & 8.54$\pm$0.05 & 7.39$\pm$0.04 & 23.02$\pm$0.09 & 0.195$\pm$0.003 \\
70CET   & 8.54$\pm$0.05 & 7.41$\pm$0.04 & 22.88$\pm$0.09 & 0.195$\pm$0.003 \\
100CET & 8.56$\pm$0.05 & 7.41$\pm$0.04 & 22.68$\pm$0.09 & 0.196$\pm$0.003 \\
\hline
\end{tabular}
\caption{The average dimensions of the simulation box and
the area per molecule of the bilayers obtained from simulation.}
\label{table:dimension}
\end{table}
\end{center}
%
%%
%\begin{center}
%\begin{table}
%\begin{tabular}{l|r|r|r|r|r}
%\hline 
%System & 0CET & 30CET & 50CET & 70CET & 100CET \\
%\hline
%$L_x$(nm) & 8.50$\pm$0.06 & 8.51$\pm$0.05 & 7.81$\pm$0.04 & 7.67$\pm$0.07 & 7.79$\pm$0.04 \\
%$L_y$(nm) & 7.38$\pm$0.05 & 7.40$\pm$0.04 & 7.76$\pm$0.04 & 7.89$\pm$0.07 & 7.79$\pm$0.04 \\
%$L_z$(nm) & 23.38$\pm$0.09 & 23.17$\pm$0.09 & 23.91$\pm$0.06 &  23.85$\pm$0.06 & 23.61$\pm$0.06 \\
%$\sigma$ (nm$^2$) & 0.186 & 0.186 &0.187 & 0.187 & 0.186  \\
%\hline
%\end{tabular}
%\caption{The average dimension of the simulation box. 
%The avareage area per molecule is very close to experimental
%results of 0.190 nm$^2$\cite{Eccleston2000}. }
%\label{table:dimension}
%\end{table}
%\end{center}
%%

\begin{figure*}[t!]
    \centering
    \begin{subfigure}[t]{0.30\textwidth}
        \centering
        \includegraphics[height=2.2in]{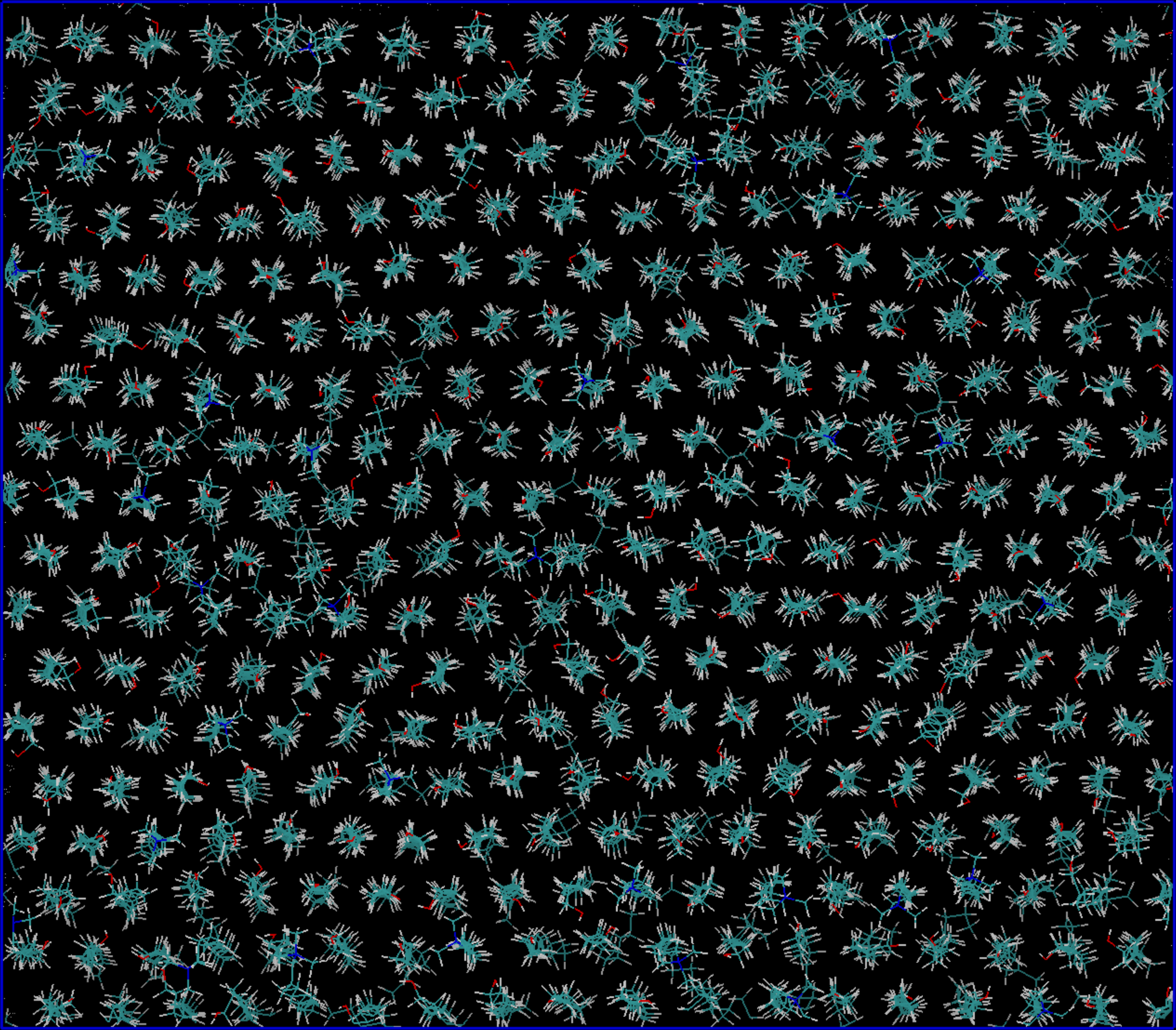}
        \caption{}
    \end{subfigure}
	~
    \begin{subfigure}[t]{0.30\textwidth}
        \centering
        \includegraphics[height=2.2in]{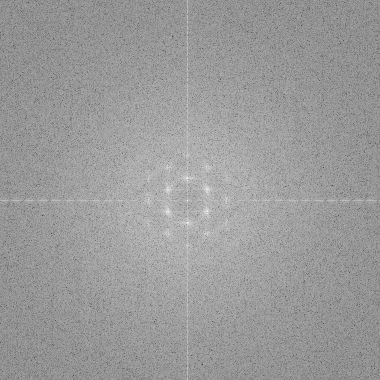}
        \caption{}
    \end{subfigure}
	~
    \begin{subfigure}[t]{0.34\textwidth}
        \centering
        \includegraphics[width=1.8in]{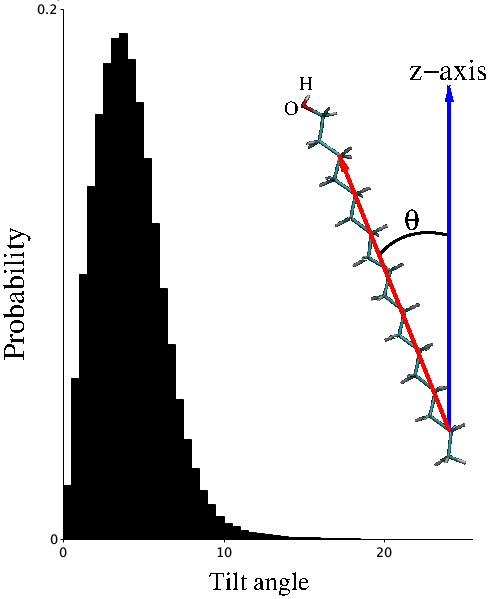}
        \caption{}
    \end{subfigure}
    \caption{(Color online) (a) Top view of the upper leaf of the bilayer 
for the 100\% stearyl system as looking down along the $z$ axis.
Each atom appears as projection onto the $xy$ plane.. 
The cetrimide molecules are drawn using thicker lines for
distinction. The configuration chosen is at 1$\mu$s of the simulation. 
(b) Fourier transform of image in (a) showing the first few orders 
of a sixth-fold symmetry. (c) Histogram of the tilt angle of the alkyl
to the normal direction of the bilayer, together with the
graphical definition of this tilt angle.
}
\label{fig:fourierImageJ}
\end{figure*}

One can also look at the local ordering in each leaf of the bilayer
from a different perspective. In Fig. \ref{fig:fourierImageJ}a, the configuration
of the molecules forming the ``upper" leaf of the bilayer for
the system of 100\% stearyl alcohol after 1 $\mu$s of simulation
is visualized by using the VMD program. A colored picture of the snapshot is taken,
and then Fourier analysis is performed by using ImageJ photo analysis program. 
The result is shown on  Fig. \ref{fig:fourierImageJ}b, where 
the first few order peaks of a six$-$fold symmetry
is clearly seen. This confirms once again the hexagonal nearest
neighbor peak in the radial distribution function.
The $\alpha-$gel and $\mathbf{L}_\beta$ crystal lamellar phases
of the alcohol bilayers are also distinguished from other phases by the fact that the hydrocarbon tail of the molecules lies parallel to the normal direction 
to the bilayer. This is indeed the case for our systems. 
In Fig. \ref{fig:fourierImageJ}, the histogram of the tilt angle of the vector
from C2 to C17 carbon atoms of the cetyl alcohol molecules
are plotted. As one can clearly sees, the hydrocarbon tail typically tilts
about 4$^{\circ}$ from the $z-$axis. This small angle shows that
our bilayer leaflets are in lamellar $\mathbf{L}_\alpha$
and $\mathbf{L}_\beta$ phases.

Further information on the system can be obtained by investigating
the electrostatic interaction in the bilayer and mobile Br$^-$ 
counterions. First, one also wants to study the distribution of
the neutral alcohol molecules around a charged cetrimide molecules.
To this end, the radial distribution function of the head group of 
the alcohol molecules (represented by the position of the oxygen atom) 
around the head group of the cetrimide molecules 
(represented by the position of the nitrogen atom) 
in the in$-$plane projection is calculated. The results for different cases 
are shown in Fig. \ref{fig:rdf_0_100}b. The lines correspond to radial distribution function
in the upper and lower leaves of the bilayer for the system with 100\% stearyl
alcohol (red and green), and for the system with 100\% cetyl
alcohol (blue and violet).

From the comparison with alkyl$-$alkyl radial distribution function, 
one can clearly see from these graphs that the charged head
group of the cetrimide molecules attracts the polar head group
of the alcohol towards it. The first peak location is smaller than that of
the alkyl first peak. The probability of finding the OH head group
of the alcohol near the cetrimide N$^+$(CH$_3$)$_3$ head group is also finite at small $r$.

Next, one looks at the distribution of the Br$^-$ counterions in the
water layer between the bilayers. 
Within standard self-consistent mean$-$field Poisson$-$Boltzmann (PB) theory of electrolytes \cite{landau2013statistical}, the distribution of the counterions 
in the water solution between two infinite similarly$-$charged parallel
planar surfaces at distance $L_w$ is given by \cite{Netz01, Netz02,Alderman1995}:
\begin{equation}
\rho(z) = \frac{ 1/ (2\pi l_B \Lambda^2)}{\cos^2[(z-L_w/2)/\Lambda]} ,
\label{eq:netz}
\end{equation}
where $z$ is the normal distance from one of the surfaces;
%and $L_z$ is the lamellar repeat distance which is equal to the
%$z-$dimension of our system; 
$\sigma_e$ is the charge density per unit area of the parallel charged surface. $l_B = e^2/4\pi\varepsilon_0\varepsilon_w k_BT$ is the so$-$called
Bjerrum length of the solution, $\varepsilon_0$ is the dielectric permeability of free space,
and $\varepsilon_w$ is the dielectric constant of the solution.
It is the length 
at which two elementary charges interact with Coulomb potential energy equal to the thermal energy, $k_BT$. 
In Eq. (\ref{eq:netz}), $\Lambda$ is a length related to 
the thickness of the water layer $L_w$ by the transcendent equation:
\beq
2\pi l_B (\sigma_e/e) \Lambda = \tan[L_w/2\Lambda] .
\label{eq:lambda}
\eeq

\begin{figure}[t!]
    \centering
    \includegraphics[height=2.2in]{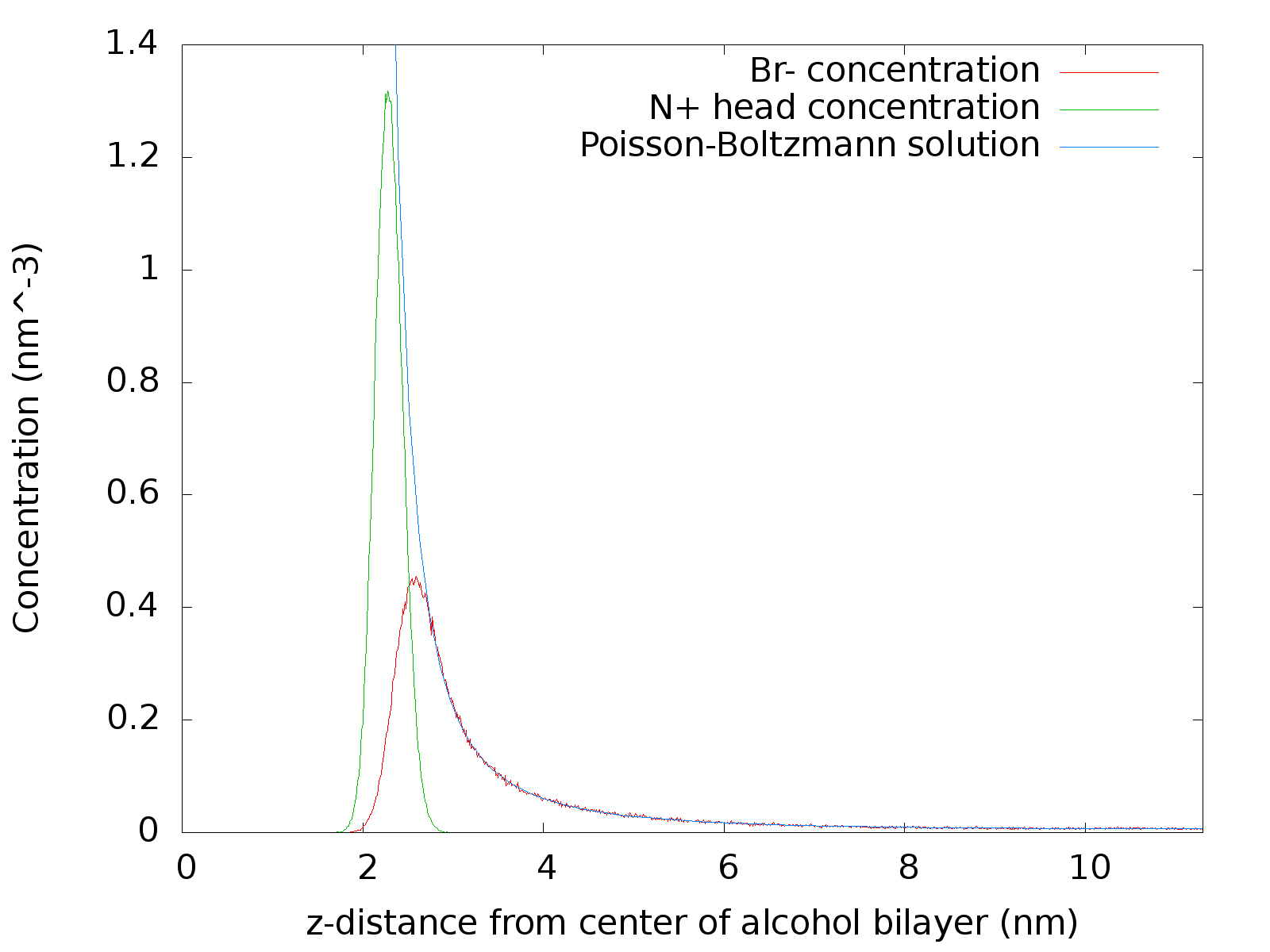}
    \caption{(Color online) The distribution of the charged
atoms in the $z-$direction perpendicular to the bilayer for
the 100CET system:
the N$^+$ atom concentration of the "upper" leaf (green),
the mobile Br$^-$ counterion concentration (red),
and the fitted Poisson$-$Boltzmann solution (blue)
showing excellent agreement in the water layer.
}
\label{fig:PBfit}
\end{figure}

In application to our bilayer system, Br$^-$ ions play the
role of mobile counterions while each leaf of the bilayer
plays the role of the charged planar surface with charge density
$$\sigma_e = 64 e /L_x L_y $$. For convenience, we measure the $z$ distance from
the center of the bilayer, then the PB solution can be rewritten as:
\begin{equation}
\rho(z) = \frac{1/(2\pi l_B \Lambda^2)}{\cos^2[(z-L_z/2)/\Lambda]} ,
\label{eq:netz2}
\end{equation}
Since the water solution is explicitly simulated in our system, by
fitting the PB solution to the Br$^-$ counterion profile, one can obtain the effective dielectric constant of the system as well as the
water layer thickness, $L_w$ and compare it to the above analysis. 
It should be noted that, due to a mismatch of the dielectric constants of the bilayer and water layer, the dielectric constant varies near the bilayer surface \cite{NetzPRL2016}. Taking into consideration the fact that the bilayer is not ideal planar surface, one expects the PB solution is applicable in about few water layers from the bilayer surface.
In this section, we chose to fit the analytic formula from a distance of
3nm from the center of the bilayer to the center of the water. 
The results of our fitting is shown in Fig. \ref{fig:PBfit} and Table \ref{table:PBEfit}.

In Fig. \ref{fig:PBfit}, typical concentrations of the positive
charges of the bilayer head group represented by the atom N$^+$,
of the negative mobile counterions Br$^-$, and the fitted PB
solution are plotted. The PB solution shows an excellent agreement to our simulated counterion profile suggesting
this mean$-$field theory is highly applicable for calculation
of electrostatics of our system.

% 0CET
% 0.00648381 +- 0.7537%
% 0.16561 +- 0.0452%
% 30CET
%a               = 0.0066759        +/- 6.17e-05     (0.9242%)
%b               = 0.166116         +/- 0.0001042    (0.06276%)
%50CET
%a               = 0.00673192       +/- 5.548e-05    (0.8242%)
%b               = 0.166243         +/- 0.0001022    (0.06146%)
%70CET
%a               = 0.0068536        +/- 5.906e-05    (0.8617%)
%b               = 0.167063         +/- 0.0001101    (0.0659%)
%100CET
%a               = 0.00682525       +/- 3.513e-05    (0.5148%)
%b               = 0.167266         +/- 7.406e-05    (0.04428%)

% new forcefield result 
\begin{center}
\begin{table}
\begin{tabular}{c||c|c|c|c|c}
\hline
System & $\Lambda$ (nm) &  $\varepsilon_w$ & $L_w$ (nm) 
& $V_w$ (nm$^3$)& $d$(nm) \\
\hline
0CET    & 6.038 & 76.85 & 18.51 & 1159.9 & 4.88 \\
30CET   & 6.019 & 75.10 & 18.44 & 1160.1 & 4.74 \\
50CET   & 6.015 & 74.59 & 18.42 & 1162.4 & 4.60 \\
70CET   & 5.896 & 73.99 & 18.32 & 1158.7 & 4.56 \\
100CET  & 5.880 & 74.02 & 18.30 & 1160.6 & 4.38 \\
\hline
\end{tabular}
\caption{Fitting parameters of PB solution, Eq. (\ref{eq:netz2})
to the Br$^-$ ion profile in the water layer from 3nm to
the center of the water layer. See text for discussion.}
\label{table:PBEfit}
\end{table}
\end{center}

Table \ref{table:PBEfit} shows the fitting results for various parameters. One sees that the ``effective" dielectric constant
of our systems is about 75, which is 10\% smaller than the 
known dielectric constant of bulk water in TIP3P model of 82
\cite{tip3pEpsilon}. Obviously, the presence of the hydrophobic hydrocarbon tail of the bilayer molecules leads to this decrease in the effective dielectric constant as one expects. Knowing $\Lambda$ and
$\varepsilon_w$, one can calculate the thickness of the water
layer $L_w$ from Eq. (\ref{eq:lambda}). 
This value is listed in column 3 of the Table
\ref{table:PBEfit}. Although, $L_w$ decreases slightly with
increasing cetyl percentage, the total water volume $V_w = L_x\times L_y\times L_w \approx 1160$nm$^3$ shown in column 4 is the same for all systems. This is because the same number of water molecules are simulated in all systems. The fitting result shows the consistency of our parameters.

Finally, knowing $L_w$, one can calculate the thickness of the bilayer
as seen electrostatically by the counterions as $d = L_z-L_w$. 
Column 5 of Table \ref{table:PBEfit} lists these values for our systems. Clearly
as the percentage of shorter cetyl molecules increases, the thickness of the bilayer decreases correspondingly. The total
decrease in bilayer thickness from 100\% stearyl system to 100\% cetyl system is 0.5 nm. Thus, the positively charged cetrimide molecules layers shrink together with the bilayer.

% Importantly, for mixed cetostearyl systems, $d\approx 4.55nm$. This shows the importance of the increase of configurational entropy to the bilayer thickness due to the tail length mismatch for these bilayers, a support
% to the prevalent experimental hypothesis.

%Notice that, although these These radial distribution also show deviations from each other at higher order
%peaks. This means that the long$-$range order in our systems is not well
%established yet within our simulation time of 1$\mu$s. This is a well$-$known limitation
%of computer simulation of bilayer membranes. The relaxation time for
%long range fluctuations in bilayer membranes is of the order of tens of microsecond
%which would goes far beyond the limit of our computer resources. Nevertheless,
%since the local order is reasonably well established in our simulation,
%many physics properties of at the short range can be investigated
%for our purpose.

\subsection{The effect of mixing ratio of cetyl and stearyl alcohol}

In this section, the effect of mixing different ratios of
cetyl and stearyl alcohol on the physical properties
of the bilayer are investigated. As we already mentioned
in the previous section, the lateral $xy$ area of the 
alcohol is very weakly dependent on the length of 
the hydrocarbon, and more on the interaction of the head groups.
However, one expects in the $z$-component, the length of the hydrocarbon would dictate the thickness of the bilayer as well as regulate the
interaction among different bilayers. This is indeed what
we observed. Table \ref{table:dimension} shows that
as the percentage of the cetyl alcohol molecules increases,
the $z$ dimension of our box decreases monotonically
from 23.38$\pm$0.09 nm to 22.68$\pm$0.09 nm. Since
the total amount of water and the pressure are the same for
all systems considered, this means the membrane thickness decreases
by 0.155 $\pm$0.03 nm per CH$_2$ group (note that
the stearyl alcohol bilayer is longer than the cetyl
alcohol bilayer by 2 carbon atoms per leaflet, or 4 carbon atoms
per bilayer). Accidentally, this is
almost the same as the typical length of a
C$-$C bond of 0.154 nm, but that does not mean
the  hydrocarbon chain is stretched because
this length also has to account for the inter-space between the leaflets.
Additionally, the shorter hydrocarbon also make
the individual molecules tilt more from the $z-$axis
making the apparent bilayer thickness decreases more.

As it is already known from the experiment data, after a few weeks,
the cetyl/stearyl mixture are always more stable than either pure or stearyl alone.
It is believed that the reason why the alcohol mixtures is more
stable is due to alkyl chain length mismatch \cite{Iwata2017, Snyder1983}. The
roughness and irregularities in the leaflets can lead to higher
configuration space for the tip of the hydrocarbon tails, 
as compared to mostly trans configuration for pure alcohol bilayers.
Of course, within computer simulation, it is an impossible task
to simulate the systems at the length scale of weeks required
for stability analysis. Nevertheless, within our micro second time scale,
one is already able to elucidate aspects of the molecular picture
of the increased configuration space of the tail tips. This
is indeed the case from our simulation results.

% new forcefield result 
\begin{center}
\begin{table*}
\begin{tabular}{||c||c|c|c|c||c|c|c|c||}
\hline
C$-$atom & 100STE& 70STE & 50STE & 30STE & 30CET & 50CET & 70CET & 100CET\\
\hline
  2 & 0.300 & 0.285 & 0.278 & 0.269 & 0.319 & 0.310 & 0.306 & 0.295\\
  3 & 0.412 & 0.399 & 0.391 & 0.384 & 0.419 & 0.411 & 0.409 & 0.403\\
  4 & 0.448 & 0.440 & 0.435 & 0.430 & 0.450 & 0.447 & 0.445 & 0.442\\
  5 & 0.462 & 0.458 & 0.455 & 0.453 & 0.463 & 0.461 & 0.460 & 0.458\\
  6 & 0.467 & 0.465 & 0.463 & 0.461 & 0.467 & 0.465 & 0.464 & 0.462\\
  7 & 0.470 & 0.467 & 0.466 & 0.465 & 0.469 & 0.467 & 0.467 & 0.465\\
  8 & 0.471 & 0.469 & 0.468 & 0.466 & 0.469 & 0.468 & 0.468 & 0.466\\
  9 & 0.471 & 0.470 & 0.468 & 0.467 & 0.470 & 0.469 & 0.468 & 0.466\\
10 & 0.471 & 0.469 & 0.468 & 0.466 & 0.470 & 0.468 & 0.467 & 0.465\\
11 & 0.470 & 0.468 & 0.467 & 0.465 & 0.468 & 0.466 & 0.466 & 0.464\\
12 & 0.468 & 0.466 & 0.464 & 0.462 & 0.465 & 0.462 & 0.462 & 0.459\\
13 & 0.463 & 0.460 & 0.458 & 0.454 & 0.458 & 0.452 & 0.453 & 0.450\\
14 & 0.454 & 0.449 & 0.447 & 0.442 & 0.446 & 0.439 & 0.438 & 0.431\\
15 & 0.436 & 0.427 & 0.420 & 0.414 & 0.418 & 0.407 & 0.404 & 0.392\\
16 & 0.414 & 0.396 & 0.385 & 0.371 & 0.383 & 0.369 & 0.363 & 0.347\\
17 & 0.375 & 0.346 & 0.331 & 0.311 & & & & \\
18 & 0.332 & 0.296 & 0.279 & 0.256 & & & & \\
\hline
\end{tabular}
\caption{Deuterium order parameter for the each methyl group
along the cetyl and stearyl alcohol molecules.}
\label{table:scd}
\end{table*}
\end{center}
%
%% old forcefield
%\begin{center}
%\begin{table*}
%\begin{tabular}{||c||c|c|c|c||c|c|c|c||}
%\hline
%C$-$atom & 100STE& 70STE & 50STE & 30STE & 30CET & 50CET & 70CET & 100CET\\
%\hline
%  2 & 0.421 & 0.415 & 0.396 & 0.387 & 0.454 & 0.450 & 0.447 & 0.441\\
%  3 & 0.452 & 0.451 & 0.442 & 0.438 & 0.473 & 0.472 & 0.472 & 0.469\\
%  4 & 0.469 & 0.469 & 0.464 & 0.462 & 0.482 & 0.481 & 0.480 & 0.479\\
%  5 & 0.478 & 0.480 & 0.476 & 0.477 & 0.485 & 0.484 & 0.484 & 0.484\\
%  6 & 0.484 & 0.485 & 0.483 & 0.483 & 0.487 & 0.486 & 0.486 & 0.486\\
%  7 & 0.486 & 0.486 & 0.486 & 0.485 & 0.487 & 0.487 & 0.487 & 0.487\\
%  8 & 0.487 & 0.487 & 0.487 & 0.487 & 0.488 & 0.487 & 0.487 & 0.487\\
%  9 & 0.487 & 0.487 & 0.487 & 0.487 & 0.488 & 0.487 & 0.487 & 0.487\\
%10 & 0.487 & 0.487 & 0.487 & 0.487 & 0.487 & 0.487 & 0.487 & 0.487\\
%11 & 0.487 & 0.487 & 0.487 & 0.487 & 0.487 & 0.487 & 0.487 & 0.487\\
%12 & 0.487 & 0.487 & 0.487 & 0.487 & 0.487 & 0.487 & 0.486 & 0.486\\
%13 & 0.487 & 0.487 & 0.487 & 0.487 & 0.486 & 0.485 & 0.485 & 0.485\\
%14 & 0.487 & 0.486 & 0.486 & 0.485 & 0.483 & 0.482 & 0.481 & 0.481\\
%15 & 0.485 & 0.485 & 0.484 & 0.483 & 0.475 & 0.471 & 0.469 & 0.468\\
%16 & 0.481 & 0.480 & 0.478 & 0.475 & 0.340 & 0.318 & 0.300 & 0.281\\
%17 & 0.469 & 0.463 & 0.456 & 0.449 & & & & \\
%18 & 0.281 & 0.259 & 0.251 & 0.231 & & & & \\
%\hline
%\end{tabular}
%\caption{Deuterium order parameter for the each methyl group
%along the cetyl and stearyl alcohol molecules.}
%\label{table:scd}
%\end{table*}
%\end{center}
%%
In order to analyze this configuration entropy of the chain tips, one calculates 
the deuterium order parameter, $-s_{CD}$. This is a sensitive measure
of the structural orientation and flexibility of the alcohol molecules in the
bilayers (for review, see Ref. \cite{VermeerSCD}). Its definition is given by
\begin{equation}
s_{CD} = \left \langle \frac{3 \cos^2\theta -1}{2} \right \rangle
\end{equation}
where the time dependent $\theta$ is the angle between the C$-$D (H atom in our case)
bond vector and a reference axis ($z-$ axis in our case). For ideal perfect crystal phase,
the C$-$H bond lies perpendicular to the $z$ axis, this order parameter is $-0.5$.
For very flexible bond, such as at very higher temperature, this bond can adopt all available 
directions and this order parameter is 0. The value $-s_{CD}$ varies in the range
0 to 0.5 dependent on the order in the system. 

On the table Table. \ref{table:scd}, the deuterium order parameter, $-s_{CD}$ for 
each methyl group along the hydrocarbon tails is listed for the longer stearyl 
and shorter cetyl alcohols.
One can obtain many important conclusions from this table.
First of all, in the middle of the chains, all the methyl groups have $s_{CD} = 0.470$
regardless of the mixing ratio. This is also the highest value of this order parameter
among all methyl groups.
Therefore, one can conclude that the middle of the hydrocarbon chain remain highly
ordered and is unaffected by the length mismatch of the alkyl chains. The role
of the middle methyl groups is to maintain
the structures of the gel or crystal phase.

The second observation one sees from this table is that the tail of the stearyl becomes
more flexible, with decreasing $s_{CD}$ order as the mixing percentage of cetyl increases. 
This trend happens for both tips of the chain (near the head group and near the
end chain). The behaviour is opposite for the cetyl alcohol molecules. As one
increases the mixing ratio of stearyl alcohol, the cetyl molecule becomes more
ordered with $-s_{CD}$ increases. Same as for stearyl alcohol, this trend happens 
to both tips of the chains. Thus, our molecular analysis of the order
parameters supports the argument of stability observed experimentally by confirming 
an increase in configuration entropy of the tips due to the length mismatch. Our results pointed out
further that this effect works on the longer chain only. The shorter chain is squeezed and 
becomes more ordered upon mixing.

The increased disorder of the longer chain, and increased order of the shorter chain also
show up when one calculates the area per molecules for individual species. This can be done
by projecting the oxygen atom of the polar head group of each alcohol and perform a voronoi
construction to calculate the area per lipid for individual molecules. The results after
averaging are listed in Table \ref{table:voronoi}. One can observe the same disorder for mixed system similar to
the deuterium order parameter analysis. For pure alcohol
systems, the standard fluctuations in the area per lipid
 for both cetyl and stearyl molecules are smaller, 0.09\AA$^2$.
However, in mixed systems, the magnitude of fluctuations
increases by 20-100\% to the values 0.11-0.19\AA$^2$. 
This
matches with the physical picture that mixed systems causes
higher configurational entropy.
Additionally,
the mean values of the area per lipid increases for
the longer stearyl molecule upon mixing, while the
mean value for the area per lipid decreases for the
shorter cetyl alcohol. This also agree with the more order
structure of cetyl alcohol in the bilayer upon mixing with
longer alcohol.
% new forcefield
\begin{center}
\begin{table}
\begin{tabular}{|c|c|c|c|c|c|c|c|c|}
\hline
System & Stearyl & Cetyl & Cetrimide \\
\hline
0CET   & 19.30$\pm$0.09  & $-$ & 19.76$\pm$0.31 \\ % all_avg 19.343 +- 0.074
30CET & 19.37$\pm$0.11 & 19.41$\pm$0.18 & 19.80$\pm$0.32 \\ % all_avg 19.422+-0.076
50CET & 19.41$\pm$0.14 & 19.48$\pm$0.14 & 19.83$\pm$0.32 \\ % all_avg 19.481+-0.079
70CET & 19.45$\pm$0.19 & 19.50$\pm$0.11 & 19.84$\pm$0.32 \\ % all_avg 19.518+-0.083
100CET & $-$ & 19.55$\pm$0.09 & 19.88$\pm$0.33 \\ % all_avg 19.579+-0.084
\hline
\end{tabular}
\caption{Aveaged area per molecule calculated on invidual molecule basis using
voronoi construction method. }
\label{table:voronoi}
\end{table}
\end{center}
%
% Original forcefield
%\begin{center}
%\begin{table}
%\begin{tabular}{|c|c|c|c|c|c|c|c|c|}
%\hline
%System & Stearyl & Cetyl & Cetrimide \\
%\hline
%0CET   & 18.62$\pm$0.09  & $-$ & 18.62$\pm$0.34 \\
%30CET & 18.67$\pm$0.11 & 18.39$\pm$0.19 & 18,57$\pm$0.32 \\
%50CET & 18.75$\pm$0.14 & 18.62$\pm$0.14 & 18.84$\pm$0.35 \\
%100CET & $-$ & 18.74$\pm$0.09 & 18.88$\pm$0.34 \\
%\hline
%\end{tabular}
%\caption{Aveaged area per molecule calculated on invidual molecule basis using
%voronoi construction method. }
%\label{table:voronoi}
%\end{table}
%\end{center}
%

% Lastly, during our simulation time of 400ns, we observe that for the 70CET system,  there are two cetrimide molecules leave the bilayer and diffuse into the
% water layer in between bilayers. For the 100CET system, in later
% stage one also observes the diffusion of one cetrimide molecule
% into the water layer. These facts suggest that the shorter alcohol
% have difficulty maintaining the charged cetrimide molecules within
% its bilayer boundary. In long time scale, this may imply the instability
% of these system.

Overall, our analysis supports the current experimental point of view that
the stability of our mixed ternary system at long time scale is due 
to an increase in the configurational entropy of the chain tips due to
length mismatch. This leads to more flexibility of the bilayer to relax
and fluctuations around equilibrium concentration. 
At the same time, we show that in the middle of
the chain, the shorter alkyl becomes stiffer.
% too much cetyl percentage cause the
% cetrimide molecule to diffuse into water solution. The optimal
% mixing ratio is around 50:50 ratio.

\section{Conclusion\label{sec:conclusion}}
In this paper, we investigate the structure of cetyl and stearyl
alcohol bilayers at different mixing ratios by using atomistic
computer simulation. Unlike typical unified atom models, 
our atomistic model allows one to study additional details in the bilayer 
such as the deuterium order parameters, area per molecule and fluctuations. 
The forcefield for our molecules was derived from 
quantum mechanical calculations and general
Amber forcefield for biomolecules.

Structurally, our results agree well with various X$-$ray experiment
data on the system. At the temperature studied of 27$^{\circ}$,
all our systems exhibit the lamellar phases, specifically
the $\alpha-$gel and $\mathbf{L}_\beta$ phases, where
the local order is hexagonal. The nearest neighbor distance
is found to be 4.69\AA\ matches the experimental result.
The area per molecule of the cetostearyl alcohol ternary system is 
numerically found to be 19.5\AA$^2$, lying within 5\% of the experiment value.
The electrostatics in the system is well described
by the mean$-$field Poisson$-$Boltzmann theory with 
effective dielectric constant reduced compared to
bulk water by 10\%.

The effect of mixing ratio on the stability of the alcohol ternary system
is studied by investigating various molecular order parameters
in our system. The experimental hypothesis for the observed stability of mixed system,
as compared to pure alcohol system, is that 
this is probably due to the increased configurational
entropy of individual molecules of the longer alkyl chains. 
Our analysis of the ordering of molecules 
in the bilayer supports this hypothesis. The deuterium order parameter
for the methyl group at the chain tips for the longer stearyl
alcohol decreases upon mixing.
Our simulation results provide additional details to this hypothesis,
showing that the shorter molecule becomes more ordered
upon mixing, leaving more available space to the longer molecule.
 The in$-$plane projection
for the area per molecule for this species also show stronger
fluctuation upon mixing as compared to pure alcohol systems.
The shorter molecules is also shown to have difficulty keeping
the cetrimide remaining in the bilayer at the simulated
molar ratio of various species. It suggests that these
system would more stable at lower water and/or cetrimide molar ratio. 

Overall, through detail atomistic simulation at one microsecond
time scale, we are able to elucidate and quantify several important atomistic
details of the mechanism for stability of ternary systems
that support current dominant hypothesis as well as the atomistic
structure of the bilayer gel.

\begin{acknowledgments}
We would like to thank the financial support of
the Vietnam National Foundation for Science and Technology
NAFOSTED grant number 104.99-2016.39.
\end{acknowledgments}

\bibliography{nttpaper}

\end{document}